\documentclass[ ras]{AGUTeX}

 \usepackage[dvips]{graphicx}

\authorrunninghead{PANCHENKO ET AL.}

\titlerunninghead{STEREO/WAVES-B antenna calibration}

\authoraddr{M. Panchenko, W. Macher, H.O. Rucker, G. Fischer, T. Oswald
Space Research Institute Austrian Academy of Sciences, Schmiedlstrasse  6, A-8042 Graz, Austria (mykhaylo.panchenko@oeaw.ac.at)}

\authoraddr{B. Cecconi, M. Maksimovic
LESIA, Observatoire de Paris, 5, place Jules Janssen, 92195 MEUDON Cedex, France.}

\begin{document}

%
%

\title{In-flight calibration of STEREO-B/WAVES antenna system}
%
%

%
%



\authors{M. Panchenko,\altaffilmark{1} W. Macher,\altaffilmark{1} H.O. Rucker,\altaffilmark{1}
          G. Fischer,\altaffilmark{1}
         T. H. Oswald,\altaffilmark{1}
         B. Cecconi,\altaffilmark{2}  M. Maksimovic,\altaffilmark{2}   }

\altaffiltext{1}{Space Research Institute, Austrian Academy of Sciences, Schmiedlstrasse  6, A-8042 Graz, Austria}
\altaffiltext{2}{LESIA, Observatoire de Paris, 5, place Jules Janssen, 92195 MEUDON Cedex, France.}

%
%


\begin{abstract}

The STEREO/WAVES experiment (SWAVES) onboard the two STEREO spacecraft (Solar
TErrestrial RElations Observatory)  launched on Oct 25, 2006, is
dedicated to the measurement of the radio spectrum at frequencies between a
few kHz and 16 MHz. The SWAVES antenna system consists of 6 m long orthogonal
monopoles designed to measure the electric component of the radio waves.
With this configuration direction finding of radio sources and polarimetry
(analysis of the polarization state) of incident radio waves is possible.
For the evaluation of the SWAVES data the receiving properties of the antennas,
distorted by the radiation coupling with the spacecraft body and other
onboard devices, have to be known accurately. In the present context,
these properties are described by the antenna effective length vectors.
We present the results of an in-flight calibration of the SWAVES antennas
using the observations of the non-thermal terrestrial Auroral Kilometric
Radiation (AKR) during STEREO roll maneuvers in an early stage of the mission.
A least squares method combined with a genetic algorithm was applied
to find the effective length vectors of the STEREO-B/WAVES antennas
in a quasi-static frequency range ($L_{antenna} \ll \lambda_{wave}$)
which fit best to the model and observed AKR intensity profiles.
The obtained results confirm the former SWAVES antenna analysis by
rheometry and numerical simulations. A final set of antenna parameters are
recommended as a basis for evaluations of the SWAVES data.

\end{abstract}

%
%

%

\begin{article}

%
%

\section{Introduction} \label{Intr}

The complex radiation coupling between the antennas aboard a spacecraft and other
electric devices or metallic  structures alters  currents induced in the electric
antennas by an incident electromagnetic wave, and, therefore, causes the distortion
of the reception properties. As result the effective length vectors, which describe the
main antenna reception properties such as directional characteristics and
effective length,  differ from the expected ones based on properties of the
"stand-alone" antennas.
Therefore, for the accurate evaluation of the data acquired by the radio instruments
onboard spacecraft the reception properties of the antennas, influenced by the
spacecraft body, have to be known precisely enough. This can be done by means of
several well developed techniques \cite[]{Macher2008}, such as rheometry
\cite[]{Rucker1996, Macher2007}, anechoic chamber measurements \cite[]{Riddle1976},
in-flight calibration \cite[]{Vogl2004, Panchenko2004, Cecconi2005a} or computer simulations
\cite[]{Oswald2009,Rucker2011pre}.

The rheometry method uses the scaled model of the spacecraft suspended in an
electrolytic tank and the antenna reception properties are determined by measuring a
response of the electric antennas to the electric field in the quasi-static frequency
range, i.e. when the radio wavelength is much greater than the antenna length
(short electric antenna) \cite[]{Rucker1996, Macher2011}.
A very powerful technique to analyze the reception properties of the spaceborne
antennas is a computer-based wire or patch-grid simulation in which the spacecraft
body and antennas are modeled as mesh of wires or patches
\cite[]{Fischer2001, Oswald2009, Sampl2011, Sampl2012}.
Numerical solutions of the electric and magnetic field integral equations by means of
numerical electromagnetic codes yield the satisfying determination of the
antenna effective length vectors and antenna radiation pattern for quasi-static
as well as for higher frequency range. These computer simulations allow to predict the
reception properties of the antennas before launch of the spacecraft.

The main idea of the in-flight calibration is the determination of the reception
properties of the antennas after spacecraft launch using the natural radio
sources such as galactic background or planetary radio emission. An in-flight
calibration procedure consist of two main parts:
1) determination of the antenna system gain including effective lengths of
the antennas as well as base and antenna capacitances, and
2) determination of the antenna directivity -- i.e. the directions of the antenna effective
lengths vectors.

The antenna gain and effective lengths can be determined using the quasi-isotropic
non-thermal galactic background as a reference radio source. This method has
been implemented to derive the absolute flux density measurements of the
Cassini/RPWS \cite[]{Zarka2004_GalBkg} and the effective length of the
STEREO/WAVES electrical antenna system \cite[]{Zaslavsky2011}.

The effective antenna length vectors of the electric antennas can be
investigated by analyzing the temporal variation of the intensity of the
radio emission emitted from the point source with known location. In this
method the spaceborne antenna system must rotate with respect to the direction
of the incident radiation. The in-flight calibration method is one of the
most reliable techniques to determine the reception properties of the antenna
because the method deals with the real observations. The methodology of this
calibration is well described in \cite{Vogl2004}. The method has successfully
been applied to derive the antenna properties of radio experiments onboard
e.g. ISEE-3 \cite[]{Fainberg1985}, Voyager \cite[]{Lecacheux1987,Wang1994},
Interball-2/Polrad \cite[]{Panchenko2004} or Cassini \cite[]{Vogl2004, Cecconi2005a}

In this paper we present the results of determining the effective length vectors
of the  antennas of the WAVES instrument onboard STEREO spacecraft.
We used the terrestrial Auroral Kilometric Radiation (AKR) - intense radio
emission from auroral regions \cite[]{Gurnett1974}, as a radio source.
The antenna effective length vectors have been determined by fitting of
the model-predicted temporal variations to the time profiles of the
AKR intensity measured during STEREO/WAVES roll maneuvers.
The obtained results are compared with antenna effective length vectors
evaluated by methods of rheometry and numerical computer simulation.

The STEREO/WAVES radio experiment is described in chapter \ref{sec2}.
The method of in-flight calibration of the SWAVES antenna is presented in section
\ref{sec3}, and the results and error analysis are given in section \ref{sec4}.
In  section \ref{sec5} we discuss the results.

\section{STEREO/WAVES antenna system} \label{sec2}

Solar TErrestrial RElations Observatory (STEREO) consists of two identical
spacecraft (STEREO-A and STEREO-B), launched on October 25, 2006. The main
goals of the STEREO mission are investigations of the three-dimensional
structure and evolution of the solar coronal mass ejections (CMEs) as
well as the study of the CMEs interaction with the Earth's magnetosphere.
The set of scientific experiments onboard STEREO provide complex measurements
of the electromagnetic and local plasma waves, as well as 3D magnetic
field components, and plasma parameters like the solar wind bulk velocity,
density and temperature. After several highly eccentric geocentrical orbits
and lunar swing-by maneuvers (January 2007) the two STEREO spacecraft split
up and were inserted into a heliocentric orbit. One of the spacecraft,
STEREO-A (Stereo Ahead) leads Earth in an orbit which is slightly
closer to the Sun, while the other spacecraft, STEREO-B (STEREO Behind)
trails Earth orbiting slightly outside the terrestrial orbit.

The SWAVES radio experiment is an interplanetary radio burst tracker that
observes the generation and propagation of the radio disturbances from the Sun
to the orbit of Earth \cite[]{Bougeret2008}. SWAVES consists of fixed, high
(HFR) and low (LFR) frequency receivers as well as of a time domain sampler
(TDS). The antenna system of the SWAVES instrument include three nearly mutually
orthogonal electric monopoles $X$, $Y$ and $Z$. Each of
the monopoles has a length of 6 meters.

The super-heterodyne swept frequency receiver HFR (High Frequency Receiver)
includes two separate parts (or "logical receivers") HFR1 and HFR2 which
cover the frequency ranges 0.125 - 2 MHz and 2-16 MHz, respectively, with
50 kHz frequency resolution. The time resolution depends on the working mode.
HFR1 can work in direction finding mode providing spectral and complex
cross-spectral power densities of incident waves. HFR1 is a dual channel receiver
which can be connected to a pair of monopoles $X$, $Y$ or $Z$ or to combinations
of monopoles and "pseudo dipoles" (monopoles $X$ or $Z$ can be paired with the $Y$
antenna). Thanks to the two parallel analysis channels, HFR1 measures instantaneously four
values: autocorrelation of each of the two antennas and the real and imaginary
part of cross-correlation between pairs of the antennas (see \cite{Bougeret2008}
for more details).

In the so-called direction finding mode (DF) the HFR1 switches between the
pairs of the antenna at each frequency step, providing quasi-instantaneous
auto and complex cross-spectra density of the wave between different antenna
combination. HFR1 is operated in two special direction findings modes:

\begin{description}
  \item \textbf{DF1} - "two antenna mode", in which two possible antenna configurations are used:

      \begin{description}

          \item  {\emph{DF1 '13'}} - one monopole $E_X$ and one dipole
            $E_{ZY}$ (dipole consist of $E_Z$ and $E_Y$ monopoles).
            The output signals in this mode are:  auto
            correlations ($A_X$, $A_{ZY}$) and cross-correlations ($\Re_{X/ZY})$
            and ($\Im_{X/ZY})$.

         \item  {\emph{DF1 '31'}} - one monopole $E_Z$ and one dipole
            $E_{XY}$ (monopoles $E_X$ and $E_Y$ are connected as a dipole).
            Operating in this mode HFR1 at each frequency provides four quasi-instantaneous
            output signals: two auto correlations ($A_Z$, $A_{XY}$) and the real
            ($\Re_{Z/XY}$) and the imaginary ($\Im_{Z/XY}$) part of the cross-correlation
            between  $E_Z$ and $E_{XY}$.

     \end{description}

   \item \textbf{DF2} - "three antenna mode", uses sequence combination of
     all monopoles $E_X$, $E_Y$, $E_Z$. This mode yields nine
     quasi-instantaneous output signals: three auto correlations
     ($A_X$, $A_Y$, $A_Z$); three real parts ($\Re_{XY}, \Re_{YZ}, \Re_{ZX}$)
     and three imaginary parts ($\Im_{XY}, \Im_{YZ}, \Im_{ZX}$) of
     cross-correlations.

\end{description}

\section{Method and data preparation} \label{sec3}

\subsection{AKR observation}

Several spacecraft roll maneuvers have been scheduled at the beginning
of the mission. During this maneuvers the spacecraft were at
$90 - 140$ $R_E$ away from the Earth.
Each maneuver lasted 10 hours and consisted of 10
consecutive rolls ($6^\circ$ per minute). In particular, STEREO-A
performed three roll maneuvers: 1st on 12/18/2006 06:50-16:50, 2nd
on 12/20/2006 07:20-17:20 and 3rd on 12/23/2006 04:30-14:30. Second
spacecraft, STEREO-B, performed only two roll maneuvers on 01/29/2007
05:10-15:10 and 01/31/2007 04:50-14:50. These spacecraft roll
maneuvers can be used to determinate the effective length vector of
each antenna.

\begin{figure}[t]
   \noindent\includegraphics[width=0.5 \textwidth]{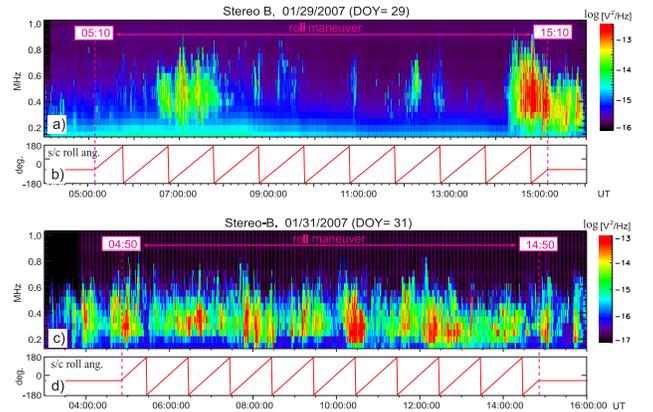}  \label{Fig1}
    \setfigurenum{1}
    \caption{ Dynamic spectra of the radio emission recorded by STEREO-B on 29 and 31 Jan., 2007 (panels a and c).
              Terrestrial AKR is observed in a frequency range up to 700-800 kHz.
              The panels b) and d) show the roll angle of the spacecraft in the GSE coordinate frame.
              This angle shows the times when STEREO-B performed roll maneuvers:
              on 29/01/2007 between 05:10-15:10 UT and 31/01/2007 between 04:50-14:50 UT.}
\end{figure}

The in-flight calibration requires the observation of the radio
emissions from a known position relative to the rotating antenna system.
During the roll maneuvers only one spacecraft, STEREO-B, observed the
terrestrial Auroral Kilometric Radiation (AKR) with the signal/noise ratio
sufficiently high to perform the antenna calibration. The reason why only
STEREO-B observed AKR is that, after the last lunar swing-by  maneuver,
STEREO-B flew by the Earth crossing the dusk side of the magnetosphere (from dayside to
nightside). Such a trajectory enabled a long-lasting (several months)
observation of the terrestrial AKR (see \cite[]{Panchenko2009}),
whose sources are known to be merely fixed in local magnetic time of the
dusk-night part of the magnetosphere. In the same time STEREO-A was
on the dayside magnetic local time sector and, therefore, it
was not able to observe the AKR.

Fig. 1 shows the dynamic radio spectra recorded by STEREO-B during
SWAVES antenna rolls on 29 and 31 Jan., 2007. The bottom panels (b) and (d)
show the roll angle of the spacecraft in the GSE (Geocentric Solar Ecliptic) coordinate frame.
The roll angle as well as the spacecraft attitude and position have been
calculated using the NASA NAIF/SPICE toolkit - a system which  provides  spacecraft ancillary
information ($http://naif.jpl.nasa.gov/naif/toolkit.html$).

As shown in  Figure 1 only in one episode, i.e.
on 31 Jan. 2007,  AKR was observed continuously during the whole roll maneuver,
whereas on 29 Jan. 2007 the radio emission was detected only during two short
time spans, i.e. 06:30-08:00 and 14:15-15:20. Therefore we have decided to
use only the observations performed on 31 Jan., 2007 between 04:50 and 14:50 UT.
During these time intervals HFR1 was operated consecutively in DF1 '13', DF1 '31'
and DF2 modes -- i.e. in  each consecutive frequency sweep  HFR1 was switched between
DF1 '13',  DF1 '31' and DF2 modes.

The first step of the data preparation deals with a subtraction of
the frequency dependent background which consists of receiver  and
galactic background noises. The level of the background has been determined
in the same way as it was done for Cassini/RPWS calibration
\cite[appendix A ]{Vogl2004}. In particular, assuming that the background
is nearly constant with time and depends only on the frequency and working
mode of the receiver the background level was determined using a histogram
of occurrence probabilities of intensities calculated for each frequency.
The histograms were obtained during time intervals when no AKR activity was detected.
Then, the background level at each frequency was defined as the intensity which
corresponds to the lower $5\%$ occurrence level (left from the  maximum of the occurrence histogram)
(see \cite{Zarka2004_GalBkg}).
The data points below the background were excluded from the analysis in order to avoid any bias
in our data set.
Since STEREO-B/WAVES was operated mainly in DF2 mode (in the beginning of the mission )
the frequency dependent background was calculated only for this
mode. The background for the DF1 mode has been assumed to be at the same level
as for DF2 mode, although, as was shown by \cite{Taubenschuss2005Th} the
background can be a slightly different for different antenna combination.
It is worth to note, that the intensity of the AKR observed on 31 Jan. 2007
was $2 - 3$ orders of magnitude higher than the averaged background level at
frequencies below 1 MHz.
Therefore, any potential inaccuracies in
background subtraction do not play a significant role in the antenna
calibration procedure.

After the background subtraction, the intensity profiles have been established
by an averaging of the data over the frequency bandwidth 225 - 475 kHz
(6 frequency channels of the HFR1 receiver). These normalized intensity profiles
are plotted as solid blue lines in Figure 3.
The standard deviation of each mean value will be used as  uncertainties of
the observations $ \sigma_i$ in equation (\ref{XiEqn}).

\subsection{Model predicted SWAVES outputs}
The open-circuit voltages on each short electrical dipole (or input
voltages at each receiver channel) can be expressed as:

\begin{equation} 
   V_i= \vec{h_i} \vec{E}
\end{equation}

where $\vec h_i$ is an effective antenna length vector, $\vec E$ is the
electric wave vector of the incoming wave and $i$ denotes the antenna,
i.e. $i={X, Y, Z}$ for DF2 mode or $i={XY,Z}$ and $i={X, ZY}$ for DF1 mode.
Each $\vec h$ vector is represented by the following spherical coordinates:
the antenna length $h_i$ and the colatitude $\zeta_i$ and the azimuth
$\xi_i$ angles. We chose the same coordinate system,  as was
used by \cite{Oswald2009} for rheometry and wire-grid calibrations of
the SWAVES antenna. In particular,  the colatitudes
$\zeta_{x,y,z}$ of each SWAVES effective antenna vectors as well as the
colatitude of the AKR source $\zeta_{s}$ are defined as an angle from
the spacecraft axis $X$ towards $Y_{sc} Z_{sc}$ plane. The azimuth angles
$\xi_{X, Y, Z}$ and $\xi_{s}$ are counted from $-Z_{sc}$
towards $Y_{sc}$.

The sample covariance of the $V_i$ in equation (1) which are actually measured by the receiver
is linearly related to the covariance of the incoming wave electric field
(which basically forms the source polarization vector).
Then, in the spacecraft coordinates  the model-predicted analytic signals
of the output voltages at each receiver can be written as \cite[]{Ladreiter1995, Vogl2004, Cecconi2005a}:

\begin{eqnarray} \label{EqMod}
\Re (\langle V_i V_j^* \rangle) & =& 0.5 S h_i h_j [(1+Q)\Omega_i\Omega_j + \\
                                &  &  U ( \Omega_i \Psi_{j} + \Omega_j \Psi_{i}) +(1-Q)\Psi_{i} \Psi_{j}  ]  \nonumber \\
\Im (\langle V_i V_j^* \rangle) &= &0.5 S h_i h_j V[\Omega_i \Psi_{j} - \Omega_j \Psi_{i}]  \nonumber
\end{eqnarray}

where  $\Omega_{i,j}= \cos \zeta_{i,j} \sin \zeta_s -  \sin \zeta_{i,j} \cos \zeta_s \cos(\phi_s - \phi_{i,j}) $,
       $\Psi_{i,j}=-\sin \zeta_{i,j} \sin(\xi_s - \xi_{i,j})$,
       $\zeta_s$ and $\xi_s$ colatitude and azimuth of the radio source in the spacecraft coordinates;
       $V_{i,j}$ -- analytical signals of the input voltages at each receiver;
       $h_{i,j}, \zeta_{i,j}, \xi_{i,j}$ - effective length, colatitude and azimuth of electric antennas;
       $S, Q, U, V$ -- Stokes parameters of the incoming wave,
       asterisk denote complex conjugate  and $\langle ... \rangle$ is a time-averaging operation.

During the spacecraft roll maneuvers SWAVES/ \\ HFR1  was
sequentially switched between DF1 '13', DF1 '31' and DF2 modes for
each measurement sweep. Therefore,  the model-predicted signals
are expressed the system of 4 equations in case of the 2 antenna
modes, DF1 '13' and DF1 '31' or 9 equations for DF2 mode. The
coefficients of these equations contain non-linear expressions of
parameters describing the the  source direction (in case of point
source approximation), Stokes parameters of the wave and the
components of the antenna effective length vectors (colatitude,
azimuth and length).

With a distant observation at $\approx 110$ $R_E$ away from Earth the AKR radio sources can be assumed as a point source.
During the roll maneuvers on 31 Jan., 2007 STEREO-B was at high latitudes over the Northern hemisphere and therefore was able
to observe only the Northern AKR sources which are mainly located in the evening sector of the magnetosphere i.e. between $\sim 18:00 - 22:00$ hours of
the magnetic local time (MLT) and  $\sim 65-75^{\circ}$ of the invariant magnetic latitude(\cite{Hanasz2003, Mutel2004}).
In our calculations we assumed that the AKR source is located at 21:00 of MLT, $70^{\circ}$ of invariant magnetic latitude and
at the altitude of $0.8 R_E$ above the Earth surface (corresponds to the sources at $\sim 300$ kHz.)
Therefore, the radio source coordinates in equation (\ref{EqMod}) are assumed as known parameters. Moreover,
using the well known AKR polarization  characteristic, i.e.  that AKR is fully circularly polarized (e.g. \cite{Panchenko2008}) with
domination of the right-handed R-X mode (\cite{Hanasz2003}) as well as  assuming that the polarization state of the AKR does not change during the
observation, we can suggest that the Stokes parameters are $Q=U=0$ and $V=-1$.
In addition, by assuming a perfect subtraction of the background contribution, one can see that the
linear system of equations (2) is homogeneous and therefore it can be normalized to the
unknown source intensity parameter, and only the antenna effective length ratio
retained.


Due to the errors in the measurements we cannot expect an exact solution of the equations
(\ref{EqMod}). Therefore the best solution in the least square
sense is sought, i.e. the sum of the squares of
the differences between the model predicted values ($P^{mod}$) and
the observations $(P^{obs})$ is minimized:
\begin{eqnarray} \label{XiEqn}
  \chi^2= \sum_{m=1}^M \sum_{n=1}^N {\sigma^{-2}_{m,n}} \left( \frac{P^{obs}_{m, n} (t_n)}{I^{obs}_{n}} - \frac{P^{mod}_{m,n} (X) }{I^{mod}_{n}} \right)^2  = min
\end{eqnarray}

where  $\mathbf{X} = \{\zeta_s, \xi_s, h_{i,j}, \zeta_{i,j},
\xi_{i,j}, S, Q, U, V \}$ are the input parameters of the model
described by equations
 (\ref{EqMod})),  index $m$ counts the output receiver channels
 (e.g. for DF2 mode:  $m=1$ denotes $A_{X}$,  $m=2$ corresponds to $A_{Y}$, etc.),
 $M$ is the total number of the receiver output channels, index $n$ denotes each data point,
 $N$ is the total number of data points in given time series and $\sigma_{n,m}$ represents uncertainties of
 each observation $P^{obs}_{n,m}$.
The normalization coefficients in equation(\ref{XiEqn}) are :
 $I^{obs,\: mod}_{n} = P^{obs,\: mod}_{X,\: n} + P^{obs,\: mod}_{Y, \: n} +P^{obs,\: mod}_{Z,\: n} $ for DF2 mode,
 $I^{obs,\: mod}_{n} = P^{obs,\: mod}_{X,\: n} + P^{obs,\: mod}_{ZY,\: n} $ for DF1 '13' mode and
 $I^{obs,\: mod}_{n} = P^{obs,\: mod}_{Z,\: n} + P^{obs,\: mod}_{XY,\: n} $ for DF1 '31'.

 Equation (\ref{XiEqn}) can be used only in assumption that the errors of observations are uncorrelated.
 This implies that the variance-covariance matrix of observations whose elements $\sigma_{m,\: n}$ are weights for the squared residuals
 is diagonal. In  fact, when the antenna system receives the strongly polarized radio signal (e.g. AKR)
 one can expect cross correlation between uncertainties of the
 measurements and as a result the non-zero off-diagonal elements in
 the variance-covariance matrix. Therefore, in more general case
 methods of the variance-covariance matrix estimation described in
\cite{Lecacheux2011} should be used. In our calculations we used the
observations with large signal/noise ratio and the expected
deviations of the electric antenna vectors from orthogonality are
not large. Therefore we assumed that the off-diagonal elements in
variance-covariance matrix are very small and we can use the
equation (\ref{XiEqn}).

The non-linear problem (\ref{XiEqn}) can be solved by means of
non-linear optimization methods. Most of these methods are based on
iterative procedures: from a starting point $x_0$ the method
produces a series of vectors $x_1, x_2, ... x_n$ which converge to
$x_{true}$ - the local minimizer for the function $\chi^2(x)$. The
main problem of such an algorithm is that the system of equations
can be ill-conditioned (numerically very close to singularity) in
certain regions of the parameters space, with no unique solution.
This is a well-known problem when the inversion is applied. In
general, the ill-conditioned system of equations describes the
situation when small fluctuations of the input data may result in a
drastic increase of errors in the solution. The other problem of the
nonlinear optimization method is the definition of the initial guess
of the model parameters. Unfortunate choosing of the initial guess
can result in finding local minima of $\chi^2(x)$ which are much
larger than its global minimum.

\subsection{Genetic algorithm (GA) }

Recently the stochastic global search techniques, such as genetic
algorithms, became a powerful tool to solve inverse problems.
The genetic algorithm (GA), the concept of which has been introduced by
\cite{Holland1975}, is a stochastic technique based on the biological
principle of survival of the fittest. The method represents each potential
solution of the optimization problem as a set of chromosomes (genome).
The algorithm begins from initialization of the initial population of
randomly selected "individuals" (potential solutions). Each "individual"
contains a set of chromosomes which represent the free parameters
of the optimization problem.
Applying the genetic operators such as mutation and crossover the fittest
"individuals" (potential solutions with the best approximation of
the problem) are determined on each iteration step.
These fittest "individuals" have a greater chance to leave most offspring
forming a next population to which the mutation and crossover operators
are applied again. In this way, a successive
approximation of the solution of the optimization problem is achieved
improving gradually. The application of the GA in astronomy and astrophysics
is well reviewed by \cite{Charbonneau1995}.

The main advantage of the GA is that this algorithm doesn't require a good initial
guess of the solution and therefore can operate in the whole space of the free
parameters of the model. Therefore, unlike "classical" nonlinear optimization methods, the GA is significantly
less sensitive to misleading local minima and in most cases the global extremum
of the function can be found. Genetic algorithms can also be
adopted to solve ill-conditioned inverse problems
\cite[e.g.][]{Mera2004}. These advantages are countered by fact that the GA requires
higher computing power. Moreover, since GA is a stochastic method which uses random
search, the resulting extremum of the optimization task is an extremum only
in probabilistic sense. In other words, the obtained results with some
probability  are only approximations of the "true" solution. The other
source of difficulties in GA is the estimation of the solution errors.
Generally, in order to obtain the resulting uncertainties the stochastic
techniques, e.g. "quick and dirty" Monte Carlo,  are used \cite[pp. 689-699]{Press1992_15_6}.

\cite{Charbonneau1995} presents the examples of solving optimization
problems by means of the GA as well as provides an open source optimization code
PIKAI which maximizes a function $f(x_1, x_2, x_3, \ldots , x_n)$ in the
n-dimensional parameter space. PIKAI has the following major
controlling parameters which determine the performance of the
algorithm: number of generations $N_g$ (or number of iterations), size of
the populations $N_p$ in each generation, mutation rate $p_m$ and crossover rate $p_c$.
 $N_g$ and  $N_p$  are most
critical parameters which determine the performance of the algorithm.
The iterative GA can be terminated either after a given number of iterations or
when some convergence criteria are satisfied, e. g. improvement
of the best solutions (the fitness level of the best individual)
falls below a given threshold.
In our case the size of the populations was $N_p=100$
and GA was terminated after $N_g=300$ iterations.
These numbers have been chosen as a reasonable compromise between computation time and
accuracy of $\chi^2$ minimization.

\section{Results and error analysis} \label{sec4}
As was mentioned in the previous section the GA can operate in the
whole space of the free parameters of the model and does
not require the initial guess of the parameters. Nevertheless, in order to
reduce the computing time, we limited the space of possible parameters in the following way:
1) tilt angles of each effective antenna vector $\vec h$ relative to
   physical antenna roods (see tables 1, 2, 3 columns "Mechanical" )
   are limited to be $30^{\circ}$,
    and 2) the effective length of each antenna $|\vec h|$ does not exceed
   $\pm 50 \%$ offset from the length of the short dipole antenna ($L/2$).

\begin{figure*} [t]
   \noindent\includegraphics[width=0.9\textwidth]{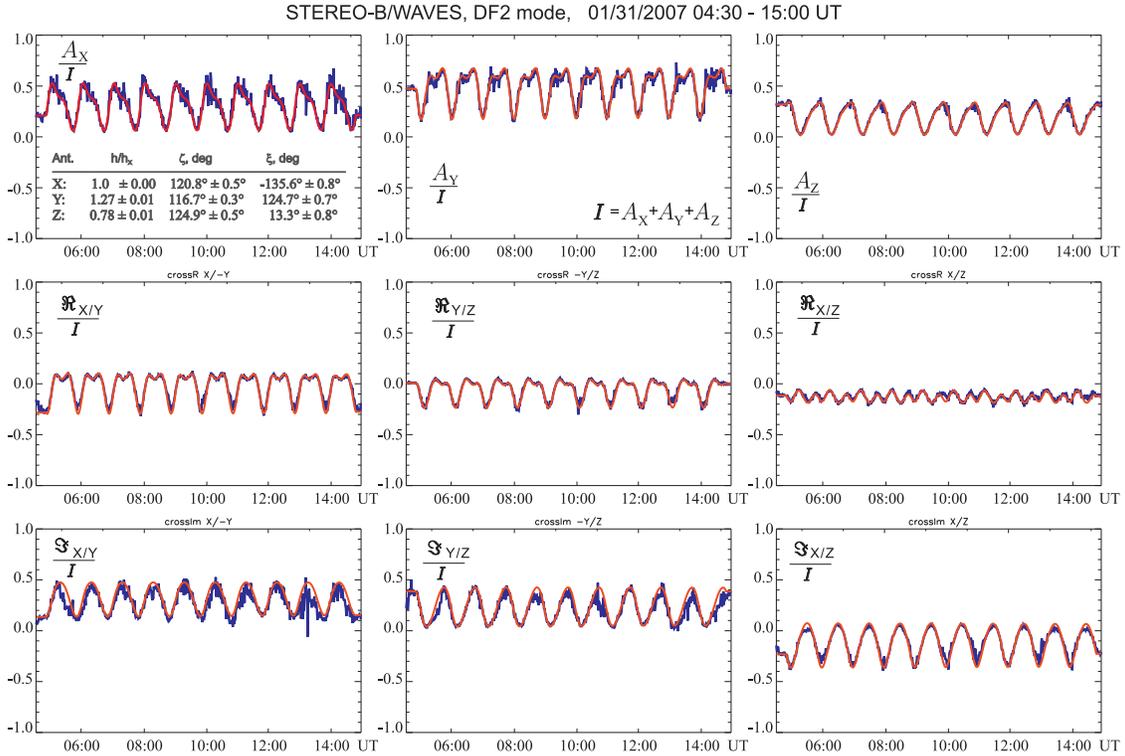}  \label{Fig3}
    \setfigurenum{2}
    \caption{ Modeled analytic signals (red lines) fitted to the measured output voltages at each channel of
              the STEREO-B HFR1 receiver working in Direction Finding (DF2) mode (blue lines).
              The modeled and measured time profiles are normalized by $I=A_X + A_Y + A_Z$.
              In DF2 mode all three antennas (X, Y, Z) are operated.
              The best fitted parameters which define the  effective antenna vectors are tabulated.}
\end{figure*}

The red lines in Figure 2 show the model predicted values for DF2 mode fitted to the
AKR observations using GA. As seen, the model  fit
the observations well.
The resulting effective antenna lengths for different antenna configuration
(optimized model parameters) are listed in tables 1, 2 and 3.

One important point is that since measurements are contaminated by
instrumental systematic errors as well as a random noise the
best fitted set of parameters is not a unique realization of the "true" parameters.
Therefore, the  errors of the fitted parameters have to be carefully estimated.
In a nonlinear least-squares methods (e.g. conjugate gradient-type techniques) the errors
can be obtained from inverse Hessian matrix of second derivatives evaluated at the best solution.
Since GA does not use the derivatives the best way is to use a Monte Carlo simulation in which
a large number of  synthetic data sets are generated from the best fitted model,
adding random noise in accordance with the measurement errors. The GA algorithm is
applied for each of these sets and finally, the standard errors of the obtained estimators are
determined.

In order to estimate the fit errors, we used a so called  "Quick and dirty" Monte Carlo or
Bootstrap method (e.g. \cite{Efron1993}) - a powerful technique
which can be applied even without information about a true underlying error distribution of the observations.
The bootstrap method generates the synthetic data sets by randomly resampling with replacing of approx 37\%
of the original data  (see for more details \cite[pp. 689-699]{Press1992_15_6}).
We synthesized M=500 bootstrapped data sets using the measured temporal variations of the AKR
intensity and then applied the  fitting procedure based on GA to each of these synthetic
data sets (each synthetic set has the same number of points as the measured AKR intensity profile). It provides 500 sets of parameters
$\mathbf{X_1}, ...,  \mathbf{X_{m}}$ each of which define the effective antenna length vectors.
Figure 3 shows the histogram of the bootstrapped parameters.
Then,  since the distribution of the bootstrapped parameters $\mathbf{X_i}$ is approximately
normal (as seen in Figure 3), the widths of the distributions give us the standard
deviations $\sigma$ of the estimated parameters.

The same calculations have been performed also for the SWAVES antennas working in DF1 mode and the
final results with one sigma errors are listed in tables 1, 2 and 3.

\section{Discussion and summary} \label{sec5}

The final set of antenna parameters of  STEREO/ \\WAVES obtained by the
in-flight calibration (for direction finding  mode DF2) are compared
with rheometry and computer simulations previously performed in Graz
and reported in \cite{Macher2007} and \cite{Oswald2009}. Table 4
summarizes the elements  of the antenna effective length vectors
using the different methods. For each method we have also checked
the goodness of the fit by calculating the reduced chi-squared:
$\xi=\chi^2 / (N*M-\nu-1) $, where $\chi^2$ is defined by equation
(\ref{XiEqn}), $N$ is the total number of observation sets, $M$ is
the number of output channels, and $\nu$ the number of fitted
parameters of the model. $\xi$ describes the discrepancy between the
observed signals of the AKR and the model predicted values, as
calculated by equation (\ref{XiEqn}) with the effective antenna
length vectors obtained by the respective method. As can be seen
from the table, the smallest $\xi$ corresponds to the in-flight
results, although the values of the goodness of the fit ($\xi$) are
very similar for all methods.

\begin{figure*} [t]
  \noindent\includegraphics[width=0.9\textwidth]{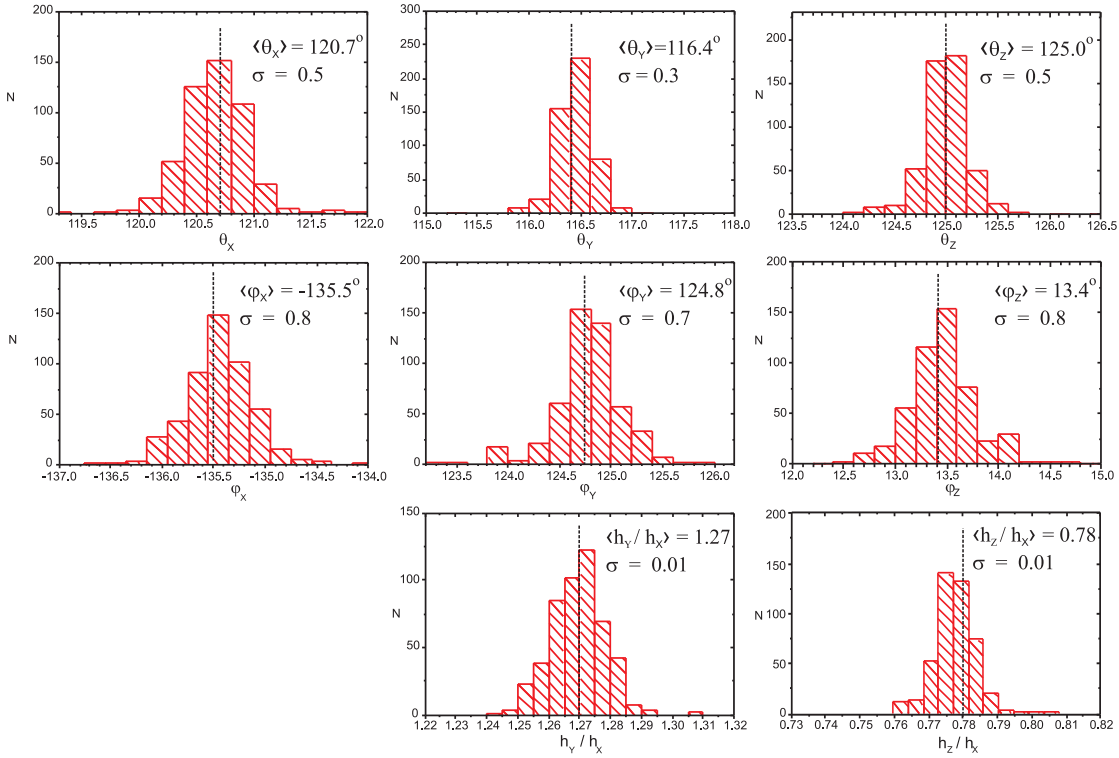}  \label{Fig4}
    \setfigurenum{3}
    \caption{ Bootstrap error analysis of best fitted parameters  $\mathbf{X}=\{h_X, \zeta_X, \xi_X; h_Y/h_X, \zeta_Y, \xi_Y; h_Z/h_X, \zeta_Z, \xi_Z\}$ which define the effective
             vectors of the SWAVES antennas  for DF2 mode.
             The histograms show the distributions of the fitted parameters obtained by application of the genetic optimization algorithm
             to M=500 synthetic data sets generated by means of the bootstrap method.
             The standard deviations of the $\mathbf{X}$
             elements are the same as the standard deviations $\sigma$  of the bootstrap histograms.
             More details are given in the penultimate paragraph of the section 4.
                 }
\end{figure*}


\begin{table*}
 \label{resDF}
\centering
\begin{tabular}{ccccccccc}
  \hline \noalign{\vskip0pt}
  & &\multicolumn{2}{c}{ Mechanical  } & & \multicolumn{3}{c} {Effective } \\
  \cline{2-4} \cline{6-8} \noalign{\vskip 0pt}
   Ant.  &  $h/h_X$ & $\zeta$  & $\xi$     &   & $h/h_X$          &  $\zeta$ & $ \xi $  &  \\
   \hline
      $X$  &  $1.0$    & $125.26^\circ$ & $-120.0^\circ$ &  & $ 1.00         $ &  $ 120.8^\circ \pm 0.5^\circ $ & $ -135.6^\circ \pm 0.8^\circ $  &  \\
      $Y$  &  $1.0$    & $125.26^\circ$ & $120.0^\circ$  &  & $ 1.27\pm 0.01 $ &  $ 116.7^\circ \pm 0.3^\circ $ & $  124.7^\circ \pm 0.7^\circ $  &  \\
      $Z$  &  $1.0$    & $125.26^\circ$ & $0.0^\circ$    &  & $ 0.78\pm 0.01 $ &  $ 124.9^\circ \pm 0.5^\circ $ & $  13.3^\circ  \pm 0.8^\circ $  &  \\ \hline \noalign{\vskip3pt}
   Ant.  &  $h/h_X$ & $\theta$  & $\varphi$     &   & $h/h_X$          &  $\theta$ & $ \varphi $  &  \\
   \hline
      $X$  &  $1.0$    & $65.9^\circ$ &  $230.8^\circ$ &  & $ 1.00         $ &  $ 52.1^\circ \pm 0.7^\circ  $ & $  229.6^\circ \pm 0.6^\circ $  &  \\
      $Y$  &  $1.0$    & $65.9^\circ$ &  $129.2^\circ$ &  & $ 1.27\pm 0.01 $ &  $ 59.4^\circ \pm 0.6^\circ  $ & $  121.5^\circ \pm 0.4^\circ $  &  \\
      $Z$  &  $1.0$    & $144.7^\circ$ & $180.0^\circ$ &  & $ 0.78\pm 0.01 $ &  $ 143.0^\circ \pm 0.5^\circ $ & $  161.8^\circ \pm 1.0^\circ $  &  \\ \hline \noalign{\vskip3pt}
\end{tabular}
\caption{Results of in-flight calibration of the SWAVES antennas
 in direction finding  mode DF2 (monopoles $X$, $Y$, and $Z$). The column headed
"Mechanical" contains relative lengths
and directions of the antenna rods. The column headed "Effective"
contains the components of the effective antenna lengths and directions with one
sigma errors.  $\zeta$ and $ \xi $ are defined in the STEREO/WAVES spherical coordinates, described in section 3.2, and
$\theta$ and  $\varphi$ are  colatitude and azimuths of the antenna vectors in the spherical coordinates related to the spacecraft main axis, i.e.
$\theta$ is measured from Z spacecraft axis, and $\varphi$ is azimuths angle from X spacecraft axis.}
\end{table*}

Also, the results of the experimental rheometry and
numerical computer simulations well agree with the in-flight calibrations
analysis within the inherent inaccuracy of the methods.
In particular, the angular components of the antenna effective length vectors
obtained by the different methods are within $3 \sigma$ errors of each other.
This confirms the high reliability of the rheometry and computer modeling for
the calibration of the antenna system.
For the effective lengths we find very small but significant differences
between the present results and those obtained by former authors
(\cite{Macher2007, Oswald2009, Zaslavsky2011}).
The most plausible explanation is that the base capacitances associated with the
antennas $X$, $Y$ and $Z$ are not exactly the same. Since base capacitances shorten the
effective length of an antenna, the results given in Table 4 indicate
$C_Y/C_X<1$ and $C_Z/C_X>1$ (the latter being confirmed by
\cite{Zaslavsky2011}, who have determined $h_X$ and $h_Z$, but not $h_Y$).

As was mentioned in section 3.1 only STEREO-B observed the AKR during the roll maneuvers and,
therefore, the in-flight calibration was applied only for the SWAVES antenna system onboard STEREO-B.
Nevertheless, since the two STEREO spacecraft are almost identical we expect that the
antenna effective length vectors of the WAVES experiment on STEREO-A are very close to
STEREO-B, determined by the in-flight calibration. This assumption was confirmed by rheometry
and computer simulations. However,  \cite{Zaslavsky2011} found slightly different results for
STEREO-A and STEREO-B by analysis of the effective antenna lengths using the galactic background.

\begin{table*}\label{resD13}
\begin{tabular}{ccccccccc}
 \hline\noalign{\vskip0pt}
 & &\multicolumn{2}{c}{ Mechanical ('13' mode)  } & & \multicolumn{3}{c} {Effective ('13' mode) } \\
 \cline{2-4} \cline{6-8} \noalign{\vskip 0pt}
   Ant.  &  $h/h_X$ & $\zeta$  & $\xi$     &   & $h/h_X$          &  $\zeta$ & $ \xi $  &  \\ 
   \hline 
      $X$   &  $1.0  $  & $125.26^\circ$ & $-120.0^\circ$ &  & $ 1.00        $ &  $ 120.4^\circ \pm 0.8^\circ $ & $ -136.3^\circ  \pm 1.1^\circ $  &  \\
      $ZY$  &  $1.414$  & $90.0^\circ$  & $150.0^\circ$  &  & $ 1.32\pm 0.01 $ &  $  92.4^\circ \pm 1.0^\circ $ & $  146.7^\circ  \pm 1.0^\circ $  &  \\ \hline \noalign{\vskip3pt}

 Ant.  &  $h/h_X$ & $\theta$  & $\varphi$     &   & $h/h_X$          &  $\theta$ & $ \varphi $  &  \\ 
      $X$   &  $1.0  $  & $65.9^\circ$ & $230.8^\circ$  &  & $ 1.00         $ &  $ 51.4^\circ \pm 0.9^\circ $ & $  229.7^\circ \pm 0.9^\circ $  &  \\
      $ZY$  &  $1.414$  & $30.0^\circ$  & $90.0^\circ$  &  & $ 1.32\pm 0.01 $ &  $ 33.4^\circ \pm 1.0^\circ $ & $  94.4^\circ  \pm 1.8^\circ $  &  \\ \hline \noalign{\vskip3pt}

\end{tabular}
\caption{Results of in-flight calibration of the SWAVES antennas in mode DF1 "13" (monopole $X$ -- dipole $ZY$).}
\end{table*}

\begin{table*}\label{resD31}
\begin{tabular}{ccccccccc}
 \hline\noalign{\vskip0pt}
 & &\multicolumn{2}{c}{ Mechanical ('31' mode)  } & & \multicolumn{3}{c} {Effective ('31' mode) } \\
 \cline{2-4} \cline{6-8} \noalign{\vskip 0pt}
   Ant.  &  $h/h_Z$ & $\zeta$  & $\xi$     &   & $h/h_Z$          &  $\zeta$ & $ \xi $  &  \\ 
   \hline \noalign{\vskip3pt}
      $XY$    &  $1.414$  & $90.0^\circ$  & $90.0^\circ$ &  & $ 1.76\pm 0.01 $ &  $  90.6^\circ \pm 0.7^\circ $ & $  88.7^\circ  \pm 0.8^\circ $  &  \\
      $Z$   &  $1.0  $  & $125.26^\circ$ & $ 0.0^\circ$ &   & $ 1.00         $ &  $ 124.3^\circ \pm 0.7^\circ $ & $ 13.8^\circ \pm 1.0^\circ $  &  \\
      \hline  \noalign{\vskip3pt}
  Ant.  &  $h/h_X$ & $\theta$  & $\varphi$     &   & $h/h_X$          &  $\theta$ & $ \varphi $  &  \\ 
    $XY$   &  $1.0  $  & $90.0^\circ$   &  $90.0^\circ$  &  & $ 1.00         $ &  $ 91.3^\circ \pm 0.8^\circ $ & $  90.6^\circ \pm 0.7^\circ $  &  \\
    $ZY$   &  $1.414$  & $144.7^\circ$  & $180.0^\circ$  &  & $ 1.33\pm 0.01 $ &  $ 143.3^\circ \pm 0.7^\circ $ & $  160.7^\circ  \pm 1.3^\circ $  &  \\ \hline \noalign{\vskip3pt}

\end{tabular}
\caption{Results of in-flight calibration of the SWAVES antennas in dipole mode DF1 "31" ( monopole $Z$ -- dipole $XY$ ).}
\end{table*}

The proper calibration of the reception properties of the spaceborne antenna system
is one of the major data processing tasks yielding an accurate evaluation of the radio
data, which is very important especially for the direction finding of  electromagnetic
waves. The in-flight calibration approach, presented in this paper, gives
us reliable results because this method uses the data acquired by the
"real" antenna-receiver configuration which are
difficult to estimate or measure during the spacecraft ground-based calibration procedures.
For example, the accurate determination of the
stray and base capacitances of the antennas plays an important role in achieving
realistic results by rheometry and computer simulation methods \cite[]{Macher2007}.

Since the AKR is observed in the frequency range below 1 MHz,
we can only study the reception properties of the antenna in
electrically short dipole approximation when the radio wavelength is much greater than the
size of the antenna system - i.e.  $L_{antenna} \ll \lambda_{wave}$.
In this frequency range (also called quasi-static frequency range) the effective
length vectors are real and constant (independent of direction).
The effective antenna length vectors on higher frequencies will be significantly
different, especially for frequencies close to the antenna resonances \cite[]{Oswald2009}.



\begin{table*}
 \label{resDF}
\centering
\begin{tabular}{ccccccc}
      $Ant.$ & $   $    &   & In-flight           & Computer simulation           & Rheometry & Mechanical  \\
    \hline
      $ $ & $h_X/h_X    $ & & $1.00             $ & $ 1.00            $ & $ 1.00            $ & $ 1.00            $  \\
      $X$ & $\zeta, deg$  & & $120.8^\circ(0.5) $ & $ 119.9^\circ  $ & $ 121.3^\circ  $ & $ 125.3^\circ  $ \\
      $ $ & $\xi, deg  $  & & $-135.6^\circ(0.8)$ & $ -135.3^\circ $ & $ -135.4^\circ $ & $ -120.0^\circ $  \\
\noalign{\vskip5pt} \cline{1-7}
      $ $ & $h_Y/h_X    $ & & $1.27(0.01)       $ & $ 1.21            $ & $ 1.26            $ & $ 1.00            $  \\
      $Y$ & $\zeta, deg$  & & $116.7^\circ(0.3) $ & $ 114.4^\circ  $ & $ 115.1^\circ  $ & $ 125.3^\circ  $ \\
      $ $ & $\xi, deg  $  & & $124.7^\circ (0.7)$ & $ 127.3^\circ  $ & $ 126.8^\circ$ & $ 120.0^\circ $  \\
\noalign{\vskip5pt} \cline{1-7}
      $ $ & $h_Z/h_X    $ & & $ 0.78(0.01)      $ & $ 0.81            $ & $ 0.84            $ & $ 1.00            $  \\
      $Z$ & $\zeta, deg$  & & $124.9^\circ(0.5) $ & $ 124.7^\circ  $ & $ 125.3^\circ  $ & $ 125.3^\circ  $ \\
      $ $ & $\xi, deg  $  & & $ 13.3^\circ (0.8)$ & $   15.5^\circ $ & $   16.4^\circ $ & $    0.0^\circ $  \\
 \noalign{\vskip5pt} \cline{1-7}
      $\xi$ & $   $ & & $ 1.08      $ & $ 1.26            $ & $ 1.25           $ & $  11.3             $  \\
%
    \hline \noalign{\vskip3pt}
\end{tabular}
\caption{Results of the STEREO-B/WAVES antenna calibration by in-flight calibration, computer simulation, and rheometry
         (computer simulation with CONCEPT and for "loaded" feeds and rheometry results from \cite{Oswald2009}). Column
         "Mechanical" contains the values which describe the relative length and direction of the antenna rods.
         The values in the parentheses denote the standard deviation of the obtained values.
         $\xi $ is the reduced chi-squared statistic  which is to check the goodness of the fit.}
\end{table*}


-----------------------------------------------------------

%

\begin{acknowledgments}

The authors are pleased to acknowledge the Plasma Physics
Data Center (CDPP) team for providing the STEREO/WAVES data.
This work was financed by the Austrian Science Fund (FWF projects P23762-N16 and P20680-N16).
In France the S/WAVES instrument has been developed with the support of both CNES and CNRS.
\end{acknowledgments}

%
%
%
%
%
%
%
%
%

\bibliographystyle{agu}

%
%

\end{article}


%
%

%
%
%
%
%
%
%


\end{document}